\documentclass[aps,prl,twocolumn,superscriptaddress]{revtex4-2}
\pdfoutput=1 
\usepackage{graphicx}
\usepackage{amsmath,mathrsfs}
\usepackage{amssymb}
\usepackage{amsfonts}
\usepackage{amsthm}
\usepackage{mathtools}
\usepackage{hyperref}
\usepackage{braket}
\usepackage{wrapfig}
\usepackage{dsfont}
\usepackage{comment}
\usepackage{lipsum}

\newcommand{\beq}{\begin{equation}}
\newcommand{\eeq}{\end{equation}}

\begin{document}

\title{Carrollian holographic duals are non-local}

\author{Jordan Cotler}
\email{jcotler@fas.harvard.edu}
\affiliation{\it Department of Physics, Harvard University, Cambridge, MA 02138, USA}
\author{Prateksh Dhivakar}
\email{pratekshd@uvic.ca}
\affiliation{\it Department of Physics and Astronomy, University of Victoria, Victoria, BC V8W 3P6, Canada}
\author{Kristan Jensen}
\email{kristanj@uvic.ca}
\affiliation{\it Department of Physics and Astronomy, University of Victoria, Victoria, BC V8W 3P6, Canada}

\begin{abstract}
Mapping the $S$-matrix of a generic theory of flat space gravity coupled to matter to correlation functions of a putative Carrollian dual, we show that bulk interactions imply boundary non-locality.
\end{abstract}

\maketitle

\emph{Introduction.}
The basic idea of the Carrollian holography program is rooted in symmetry. In an asymptotically flat spacetime, Poincar\'e transformations act on null infinity as a conformal extension of Carroll symmetry~\cite{Barnich:2006av,Barnich:2010eb,Bagchi:2010zz,Duval:2014uva}, motivating the possibility that theories of flat space gravity are dual to conformal Carrollian theories~\cite{Bagchi:2012cy,Bagchi:2016bcd}. (There is a related proposal of a celestial dual living at spatial infinity~\cite{Strominger:2017zoo,Pasterski:2021rjz,Raclariu:2021zjz}.) The pursuit of this idea has met some successes and challenges (see \cite{Bagchi:2025vri} for a recent review with an exhaustive list of references). One success is a reinterpretation of the flat space limit of AdS/CFT~\cite{Giddings:1999jq,Gary:2009ae,Penedones:2010ue,Fitzpatrick:2011hu,Raju:2012zr,Hijano:2019qmi} as a double-scaling limit, in which one simultaneously takes the large $N$ parameter of a dual CFT to infinity while also taking the boundary speed of light to zero~\cite{Bagchi:2023fbj,Alday:2024yyj,Lipstein:2025jfj}. However there are also challenges, in that the quantum mechanical behaviour of Carrollian field theories is markedly different from that inferred from flat space gravity~\cite{Basu:2018dub,Bagchi:2019xfx,Bagchi:2019clu,Banerjee:2020qjj,Henneaux:2021yzg,Chen:2021xkw,deBoer:2021jej,Hao:2021urq,Bergshoeff:2022eog,Bagchi:2022eav,Saha:2022gjw,Liu:2022mne,deBoer:2023fnj,Banerjee:2023jpi,Ecker:2024czx,Cotler:2024xhb,Cotler:2024cia,Sharma:2025rug,Cotler:2025dau}.

The standard approach in Carrollian holography so far has been to use an analogue~\cite{Bagchi:2022emh,Donnay:2022wvx,Bagchi:2023cen,Alday:2024yyj,Kraus:2024gso,Jain:2023fxc} of the AdS/CFT dictionary to map the $S$-matrix to functions. On account of the bulk Poincar\'e symmetry, these functions obey conformal Carrollian Ward identities and are interpreted as correlation functions of operators in a dual description. Much work has been done in this vein, inferring properties of putative duals from the $S$-matrix of bulk theories. So far there has been rather little work done concerning the operator interpretation of these functions. Besides studies of the quantization of Carrollian field theories, we note the work of~\cite{Banerjee:2020kaa,Nguyen:2025sqk} which studied aspects of an operator product expansion.

In this paper we are interested in the operator interpretation of putative Carrollian duals, in particular the rather basic question of whether these duals are local or not. In relativistic field theory, the definition of locality is that commutators of local operators vanish outside the light-cone \cite{Haag:1963dh}, which in turn implies causality. In Carrollian field theories, the situation is more fragile since light-cones collapse and different spatial locations are causally disconnected. Locality then mandates that all commutators of local operators vanish at nonzero spatial separation. We would like to see if this stringent condition is met.

To answer this question we must first tackle the problem of boundary operator ordering. The standard dictionary relating the $S$-matrix to Carrollian correlators \cite{Bagchi:2022emh,Donnay:2022wvx,Bagchi:2023cen,Alday:2024yyj,Kraus:2024gso,Jain:2023fxc} borrows from the Hilbert space and operators of the flat space theory, implicitly building a Hilbert space by acting with creation/annihilation operators for massless fields on the bulk vacuum. The basic boundary operators are the integral transform (in null boundary time) of those creation/annihilation operators. In this dictionary operators fall into one of two types, future and past, respectively superpositions of positive (negative)-frequency modes built from annihilation (creation) operators on future (past) null infinity. Future operators are of course related to past ones by the $S$-matrix. Bulk causality implies that future operators commute with each other, and past operators commute with each other. In free theories, future operators and past operators commute unless they have the same spatial momentum, which in a `dual' means they commute at nonzero spatial separation. However, unsurprisingly, we find that this is no longer true in the presence of bulk interactions, implying a violation of boundary locality. This is the main result of our paper.

One might wonder if the existing dictionary mapping the $S$-matrix to Carrollian correlators is the culprit, and that this work implies the need for an improved version. While in our opinion the current dictionary is not satisfactory, since operator insertions are built from superpositions of only positive or negative frequency modes, the central physical fact underlying our analysis -- that bulk interactions imply in-operators no longer commute with out-operators -- will appear in any improved dictionary and we expect will lead to the same result. Indeed, we attempt to construct a modified dictionary that treats past and future equally, and yet under it, bulk interactions still imply boundary non-locality.

Our result can also be understood intuitively using the flat-space limit of AdS/CFT~\cite{Penedones:2010ue,Fitzpatrick:2011hu,Hijano:2019qmi}. There, past operators are really CFT operators acting in a small time band concentrated around a fixed boundary time while future operators live in a small time band a light-crossing time $\Delta \tau=L \pi$ later, with $L$ the AdS radius. From this point of view future operators live in the light-cone of past operators and so, of course, they do not commute. 

We explore these ideas using the simplest possible bulk model, scalar $\phi^3$ theory in four dimensions, but as we have stressed above, the underlying physics is general. 

While known Carrollian quantum theories behave dramatically differently from what we infer from the flat space $S$-matrix, one might hope that there are yet unknown Carrollian duals to flat space gravity that exist as solutions to a suitable bootstrap. However, the absence of locality in a Carrollian dual description implies the lack of an operator product expansion, the most important ingredient in any bootstrap, and therefore dashes this hope upon a rock.

Our focus here is exclusively with the possibility of a Carrollian dual to the flat space $S$-matrix, but it is worth noting that putative celestial duals are also non-local, although for different reasons~\cite{Freidel:2021dfs,Ball:2022bgg,Banerjee:2022wht,Fiorucci:2023lpb}.

The remainder of this paper is organized as follows. In the next Section, we review the existing dictionary between the $S$-matrix of massless fields and conformal Carrollian correlators. We then demonstrate our central result, that bulk interactions mandate boundary non-locality. We go on to construct an improved dictionary, but this improvement does not restore locality. In the Appendix, we discuss other regulatory schemes for local Carrollian theories and why old computations of ``entanglement entropy'' in a putative Carrollian dual to 3d gravity coupled to matter cannot be interpreted as a measure of entanglement.

\emph{Review.}
We briefly recall the map between $S$-matrices in asymptotically flat spacetimes and Carrollian correlation functions~\cite{Bagchi:2022emh,Donnay:2022aba,Donnay:2022wvx}. There is a similar dictionary relevant for ``celestial holography.'' For definiteness, consider a real scalar $\phi$ in $3+1$-dimensional Minkowski space. In the free theory one has the standard mode expansion
\begin{equation}
\phi(x) = \int \frac{d^3 p}{(2\pi)^3} \frac{1}{\sqrt{2 E_{\mathbf p}}} \Big( a(\mathbf p)\, e^{-i p\cdot x} + a^\dagger(\mathbf p)\, e^{i p\cdot x} \Big)\,.
\end{equation}
With local interactions, the asymptotic field is expressed in terms of $a_{\text{in/out}}$ operators. Since our interest is massless scattering, we use the standard null-momentum parametrization~\cite{Pasterski:2016qvg}
\begin{equation}\label{eq:para}
 p^\mu = \omega\, \big( 1+z\bar z,\, z+\bar z,\, -i(z-\bar z),\, 1 - z\bar z \big)\,,
\end{equation}
where $\omega>0$ is the energy and $(z,\bar z)$ are stereographic coordinates on the celestial sphere.

In the conformal compactification, $\mathscr I^+$ is coordinatized by $(u,z,\bar z)$ and $\mathscr I^-$ by $(v,z,\bar z)$. The map from asymptotic operators to Carrollian ones is implemented by a Mellin-like transform in null time \cite{Banerjee:2018gce,Banerjee:2019prz,Bagchi:2022emh,Donnay:2022aba,Donnay:2022wvx}:
\begin{equation}\label{eq:mapfields}
\begin{aligned}
\Phi_{\text{out}}^{\Delta}(u,\!z,\!\bar z) &= \int_0^\infty d\omega\, \omega^{\Delta-1} \Big( a_{\text{out}}(\omega,\!z,\!\bar z)\, e^{-i\omega u}+ (\text{h.c.})\Big)\,,\\
\Phi_{\text{in}}^{\Delta}(v,\!z,\!\bar z) &= \int_0^\infty d\omega\, \omega^{\Delta-1} \Big( a_{\text{in}}(\omega,\!z,\!\bar z)\, e^{-i\omega v} + (\text{h.c.})\Big)\,,
\end{aligned}
\end{equation}
where we keep the conformal dimension $\Delta$ arbitrary but for clarity, we have not included the arbitrary normalization constant. This flexibility captures multi-particle states~\cite{Kulp:2024scx} and simultaneously avoids the IR issues that afflict correlators where we assign the naive dimension $\Delta=1$ to the dual of massless fields~\cite{Bagchi:2023cen}. Indeed, \cite{Kulp:2024scx} emphasizes that multi-trace primaries in the parent CFT generically lead to $\Delta$ differing from the extrapolate dictionary of \cite{Donnay:2022wvx}.

The BMS$_4$ algebra~\cite{Bondi1962,Sachs1962} takes the form
\begin{align}
\nonumber
[L_n,L_m] &= (n-m)\, L_{n+m}\,, \,\,\, [\bar L_n,\bar L_m] = (n-m)\, \bar L_{n+m}\,, \\
\label{eq:bms}
[L_n,M_{r,s}] &= \Big(\tfrac{n+1}{2} - r\Big)\, M_{r+n,s}\,, 
\\
\nonumber
 [\bar L_n,M_{r,s}] &= \Big(\tfrac{n+1}{2} - s\Big)\, M_{r,s+n}\,, \quad [M_{r,s},M_{p,q}] = 0\,.
\end{align}
Here $\{L_n,\bar L_n\}$ generate superrotations on the celestial sphere and $M_{r,s}$ generate supertranslations. The inspiration behind the Carrollian approach to flat-space holography rests on the isomorphism between~\eqref{eq:bms} and the infinite-dimensional lift of the $3d$ conformal Carrollian algebra~\cite{Bagchi:2010zz,Duval:2014uva}. Using the transformation properties of the creation/annihilation operators, one finds that the boundary fields in \eqref{eq:mapfields} transform as BMS$_4$ primaries \cite{Banerjee:2018gce}. Thus \eqref{eq:mapfields} effects a change of basis from plane waves to conformal primaries \cite{Pasterski:2017kqt,Banerjee:2018gce,Bagchi:2022emh,Donnay:2022aba,Donnay:2022wvx}.

Inserting~\eqref{eq:mapfields} into $S$-matrix elements, specifically matrix elements of products of in/out operators, gives a Mellin integral representation of celestial correlators. For a process with $p$ outgoing and $n-p$ incoming particles we have
\begin{widetext}
\begin{align}
\begin{split}
\label{eq:themap}
 &\big\langle \Phi_{\text{out}}^{\Delta_1}(u_1,z_1,\bar z_1)\cdots \Phi_{\text{out}}^{\Delta_p}(u_p,z_p,\bar z_p)\Phi_{\text{in}}^{\Delta_{p+1}}(v_{p+1},z_{p+1},\bar z_{p+1})\cdots \Phi_{\text{in}}^{\Delta_n}(v_n,z_n,\bar z_n) \big\rangle 
 \\
&\quad= \prod_{i=1}^{p} \int_0^\infty d\omega_i\, \omega_i^{\Delta_i-1}\,e^{-i\omega_i (u_i-i\epsilon)}\, 
\prod_{j=p+1}^{n} \int_0^\infty d\omega_j\, \omega_j^{\Delta_j-1}\,e^{i\omega_j (v_j+i \epsilon)}S(\omega_1,z_1,\bar z_1;\ldots\,,\omega_n,z_n,\bar z_n)\,,
\end{split}
\end{align}
\end{widetext}
with $\epsilon>0$ implementing the usual $i\epsilon$ prescription to render the Mellin integrals convergent~\footnote{Only the combinations $a_{\text{out}}..a_{\text{out}}\,a_{\text{in}}^\dagger..a_{\text{in}}^{\dagger}$ contribute inside $S$-matrix elements; terms with $a_{\text{out}}^\dagger$ on the left or $a_{\text{in}}$ on the right annihilate the in/out vacua and drop out prior to the Mellin transform.}. 

Using the (antipodal) identification of the celestial spheres at $\mathscr I^+$ and $\mathscr I^-$ \cite{Donnay:2022wvx}, one may trade $v$ for $u$ and work entirely at $\mathscr I^+$. The dynamical content of scattering is then encoded in the ``out'' sector and, correspondingly, in the resulting Carrollian correlators. 

\newpage
\emph{Boundary non-locality.}
Consider massless scalar $\phi^3$ theory. In perturbation theory, the interacting theory has an interacting vacuum $| \Omega \rangle$ evolved from the free field vacuum $|0 \rangle$ using the interaction picture Hamiltonian. Using standard QFT techniques to express $a_{\text{out}}(\mathbf{p})$ in terms of $a_{\text{in}}(\mathbf{p})$, we find~\footnote{We follow the standard notation where boldface symbols of the form $\mathbf{p}_i$ are used to denote vectors in $3d$ space, whereas four momenta are denoted by $p_i$.}
\begin{align}
\begin{split}\label{eq:4ptcomm}
        \langle \Omega | &a_{\text{out}}(\mathbf{p}_1) \, [a_{\text{out}}(\mathbf{p}_2), a^{\dagger}_{\text{in}}(\mathbf{p}_3)] \, a^{\dagger}_{\text{in}}(\mathbf{p}_4) | \Omega \rangle
        \\
        & \qquad  = \mathcal{A}_{\rm free} + \mathcal{A}_{\rm exchange} + \mathcal{A}_{\rm collinear} + O(\lambda^3)\,,
\end{split}
\end{align}
where
\begin{align}
       \nonumber
               &\mathcal{A}_{\text{free}} = (2\pi)^6 (2 E_{\mathbf{p}_1}) (2 E_{\mathbf{p}_2}) \delta^{(3)}(\mathbf{p}_1 - \mathbf{p}_4) \delta^{(3)}(\mathbf{p}_2 - \mathbf{p}_3) \, , \\
               \label{eq:commres}
               &\mathcal{A}_{\text{exchange}} =  i(2\pi)^4 \lambda^2 \, \delta^{(4)}(p_1+p_2-p_3-p_4) 
               \\
               \nonumber
               & \qquad \qquad \qquad \times \left[ \dfrac{1}{s+i \varepsilon} + \dfrac{1}{t+i \varepsilon} + \dfrac{1}{u+i \varepsilon} \right] \, , 
       \end{align}
where $s =(p_1+p_2)^2$, $t = (p_1-p_3)^2$, $u = (p_1-p_4)^2$, and there is a $O(\lambda^2)$ collinear contribution $\mathcal{A}_{\rm collinear}$ delta-function supported on $p_2\propto p_3$ or $p_2 \propto p_4$ whose details depend on the renormalization scheme.

Here $\mathcal{A}_{\rm exchange}$ is the standard sum of exchange diagrams contributing to the $S$-matrix and is the one responsible for boundary non-locality. It comes solely from the time-ordered part of~\eqref{eq:4ptcomm}. There is no such contribution in the out-of-time-ordered term $\langle \Omega | a_{\text{out}}(\mathbf{p}_1) a^{\dagger}_{\text{in}}(\mathbf{p}_3) a_{\text{out}}(\mathbf{p}_2)  a^{\dagger}_{\text{in}}(\mathbf{p}_4) | \Omega \rangle$.  To see this, note that at the level of asymptotic states, we have $\hat{S}\,a_{\text{in}}^\dagger(p)\,|\Omega\rangle = a_{\text{out}}^\dagger(p)\,|\Omega\rangle + [\text{multi-particle states}]$~\footnote{$\hat{S}$ is the evolution operator generated by the interaction picture Hamiltonian.}.  In massless $\lambda \phi^3$ theory, the first subleading terms at $\mathcal{O}(\lambda)$ come from $1\to 2$ splittings which are kinematically forced to be collinear when all legs are on-shell.  In the time-ordered matrix element $\langle \Omega | a_{\text{out}}(\mathbf{p}_1) a_{\text{out}}(\mathbf{p}_2) a^{\dagger}_{\text{in}}(\mathbf{p}_3) a^{\dagger}_{\text{in}}(\mathbf{p}_4) | \Omega \rangle$, each $a_{\text{in}}^\dagger$ is converted by $\hat{S}$ into an $a_{\text{out}}^\dagger$ plus its collinear multi-particle tail, and the ${\cal O}(\lambda^2)$ piece in which the two resulting one-particle out-states scatter through the usual $2 \to 2$ kernel produces $\mathcal{A}_{\text{exchange}}$.  By contrast, in the mixed ordering $\langle \Omega | a_{\text{out}}(\mathbf{p}_1) a^{\dagger}_{\text{in}}(\mathbf{p}_3) a_{\text{out}}(\mathbf{p}_2)  a^{\dagger}_{\text{in}}(\mathbf{p}_4) | \Omega \rangle$ each $a_{\text{out}}(\mathbf{p}_i)$ is already adjacent to an $a_{\text{in}}^\dagger(\mathbf{p}_j)$, so at generic kinematics the only contribution at $\mathcal{O}(\lambda^2)$ comes from the diagonal $1\to 1$ part $\hat{S}\,a_{\text{in}}^\dagger(p)\,|\Omega\rangle \sim a_{\text{out}}^\dagger(p)\,|\Omega\rangle$, which enforces $\mathbf{p}_1 = \mathbf{p}_3$ and $\mathbf{p}_2 = \mathbf{p}_4$.  The multi-particle components in $\hat{S}\,a_{\text{in}}^\dagger(p)\,|\Omega\rangle$ can contribute only when the extra on-shell quanta are exactly collinear with one of the external legs; these degenerate configurations are precisely isolated by the delta functions in $\mathcal{A}_{\text{collinear}}$, and so do not build up an additional exchange kernel.

It is useful to record the commutator
\begin{widetext}
\begin{equation}\label{eq:outincomm}
        \begin{split}
             \left[ a_{\text{out}}(\mathbf{p}_1), a^{\dagger}_{\text{in}}(\mathbf{p}_2)  \right] &= (2\pi)^3 2 E_{\mathbf{p}_1} \, \delta^{(3)}(\mathbf{p}_1 - \mathbf{p}_2) + i \lambda \int d^4x \, e^{i(p_1-p_2) \cdot x} \phi_I(x) 
             \\
             & \hspace{-.5in}+ \dfrac{\lambda^2}{12} \int \int d^4x \, d^4x' \, e^{i p_1 \cdot x} \, \theta(x^0-x'^0) \left( - 2 e^{-i p_2 \cdot x} \left[ \phi_I(x) , \phi^3_I(x') \right] + 6 e^{-ip_2 \cdot x'} \left[ \phi^2_I(x) , \phi^2_I(x') \right]  \right) +O(\lambda^3)\, ,
        \end{split}
\end{equation}
\end{widetext}
where $\phi_I$ is the interaction picture field. It is these $\phi_I$ dependent terms that imply boundary non-locality.

From the standard dictionary \eqref{eq:themap}, an expression of the form \eqref{eq:4ptcomm} corresponds to
\begin{widetext}
\begin{equation}\label{eq:commbdy}
          \begin{split}
               \langle \Phi^{\Delta_1}_{\text{out}}(u_1,z_1,\bar{z}_1) &\left[ \Phi^{\Delta_2}_{\text{out}}(u_2,z_2,\bar{z}_2) , \Phi^{\Delta_3}_{\text{in}}(u_3,z_3,\bar{z}_3) \right] \Phi^{\Delta_4}_{\text{in}}(u_4,z_4,\bar{z}_4) \rangle
               \\
                & =  \int \left( \prod_{i=1}^{4} d\omega_i \, \omega_i^{\Delta_i-1} e^{i \epsilon_i \omega_i u_i} \right) \langle \Omega | a_{\text{out}}(\mathbf{p}_1) \, [a_{\text{out}}(\mathbf{p}_2), a^{\dagger}_{\text{in}}(\mathbf{p}_3)] \, a^{\dagger}_{\text{in}}(\mathbf{p}_4) | \Omega \rangle \, .
          \end{split}
\end{equation}
\end{widetext}
From the boundary perspective, one would na\"ively expect this expression to break down into a product of two point structures weighted by appropriate space-time dependent factors. Indeed according to the standard HPPS argument \cite{Heemskerk:2009pn}, bulk locality in AdS arises from the solutions to the crossing conditions on the CFT 4-point correlator.  One would obtain those crossing conditions via an OPE expansion and equating the resulting OPE blocks. If one is drawing analogy from AdS/CFT, one would expect \eqref{eq:commbdy} would break down into a product of Carrollian OPEs \cite{Banerjee:2020kaa,Nguyen:2025sqk} weighted by appropriate OPE blocks. This expectation is spectacularly wrong and \eqref{eq:commbdy} is inconsistent with an OPE on account of the exchange term.

The dictionary \eqref{eq:themap} is known to give the complicated result~\cite{Banerjee:2019prz,Bagchi:2023cen,Mason:2023mti,Alday:2024yyj}
\begin{widetext}
\begin{equation}\label{eq:extransf}
        \begin{split}
             & \int \left( \prod_{i=1}^{4} d\omega_i \omega_i^{\Delta_i-1} e^{i \epsilon_i \omega_i u_i} \right) \mathcal{A}_{\text{exchange}} =\dfrac{\lambda^2}{4}  \Big[ \dfrac{\Gamma(\Delta_1+\Delta_2+\Delta_3+\Delta_4-6)}{z^2_{13} \bar{z}^2_{13} z_{24} \bar{z}_{24} } \dfrac{(\sigma^*_1)^{\Delta_1-2} (\sigma^*_2)^{\Delta_2-1} (\sigma^*_3)^{\Delta_3-2} (\sigma^*_4)^{\Delta_4-1} }{[i(\sigma^*_1 u_1 + \sigma^*_3 u_3 - \sigma^*_2 u_2 - \sigma^*_4 u_4)]^{\Delta_1+\Delta_2+\Delta_3+\Delta_4-6}}  \\
               &\qquad \qquad \qquad \qquad \qquad \qquad \qquad \qquad \qquad \times \prod_{i=1}^4 \mathbb{I}_{[0,1]}(\sigma^*_i) + \text{($s$-channel)} + \text{($u$-channel)} \Big] \delta(|z-\bar{z}|) \, ,
        \end{split}
\end{equation}
\end{widetext}
where
\begin{equation}
        \begin{split}
               \sigma^*_1 = \dfrac{1}{D} \dfrac{z_{24} \bar{z}_{34} }{ z_{12} \bar{z}_{13}}\,,  ~~~ \sigma^*_2 &= \dfrac{1}{D} \dfrac{z_{34} \bar{z}_{14} }{ z_{23} \bar{z}_{12}} \, ,  ~~~ \sigma^*_3 = \dfrac{1}{D} \dfrac{z_{24} \bar{z}_{14}}{ z_{23} \bar{z}_{13} }\,,  \\
              \sigma^*_4 = \dfrac{1}{D} \, ,~~~ D &= 2 \dfrac{z_{24} \bar{z}_{34} }{z_{12} \bar{z}_{13} } + 2 \dfrac{z_{24} \bar{z}_{14}}{z_{23} \bar{z}_{13}}\,,
          \end{split}
\end{equation}
and $z,\bar{z}$ are the conformal cross ratios
\begin{equation}
         z = \dfrac{z_{12} z_{34}}{z_{13} z_{24}}\, , ~~~~ \bar{z} = \dfrac{\bar{z}_{12} \bar{z}_{34}}{\bar{z}_{13} \bar{z}_{24}}\,.
\end{equation}
$\mathbb{I}_{[0,1]}(x)$ is the indicator function
\begin{equation}
           \mathbb{I}_{[0,1]}(x) = \begin{cases}
			1, & \text{if $x \in$ [0,1]}\\
            0, & \text{otherwise}
		 \end{cases}.
\end{equation}
The role of the indicator function is simply to constrain the magnitude of the conformal cross ratio depending on the particular channel under consideration \cite{Pasterski:2017ylz}. By choosing $\Delta_i \neq 1$, we avoided IR singularities in \eqref{eq:extransf}. For the $t$-channel process, the indicator functions imply $0<z<1$. To avoid clutter we have only indicated the $t$-channel process. The others have analogous structures but most importantly are proportional to $\delta(|z-\bar{z}|)$. 

The delta function $\delta(|z - \bar{z}|)$ cannot arise from an ultra-local Carrollian OPE like that appearing in the free field result. Rather it enforces the constraint that the conformal cross ratio is real. We can understand this from the following fact. Using the $SL(2;\mathbb{C})$ invariance to move three of the insertions to the equator of the celestial sphere, momentum conservation enforces that the fourth point lies on the equator as well. This is a non-local structure that depends on all four insertions in the boundary field theory. It has no counterpart in the standard way AdS bulk locality arises from the solutions to the crossing equation of the boundary CFT four-point function \cite{Heemskerk:2009pn}. Instead, in flat space, local bulk interactions generate non-local structures in a Carrollian dual.

Let us ask if, as in ordinary field theory, analytic continuation in boundary time relates different orderings to each other. Consider the free theory. Since $\langle 0|a^{\dagger}_{\text{in}}(\mathbf{p}_2)a_{\text{out}}(\mathbf{p}_1)|0\rangle = 0$, we infer
\begin{align}\label{eq:commoutin}
\begin{split}
&\langle [\Phi_{\rm out}^{\Delta}(u_1,z_1,\bar{z}_1),\Phi_{\rm in}^{\Delta}(u_2,z_2,\bar{z}_2)]\rangle \\
& \qquad \qquad \quad = \langle \Phi_{\rm out}^{\Delta}(u_1,z_1,\bar{z}_1)\Phi_{\rm in}^{\Delta}(u_2,z_2,\bar{z}_2)\rangle \,,
\end{split}
\end{align}
which is \emph{not} what one infers from analytic continuation of the RHS in boundary time. 

For this reason we obtain out-of-time-ordered correlators not by analytically continuing in boundary time but by directly considering the matrix elements of out-of-time-ordered strings of $a_{\rm out}$'s and $a_{\rm in}^{\dagger}$'s.

\begin{figure}[t]
\begin{center}
\includegraphics[scale=.35]{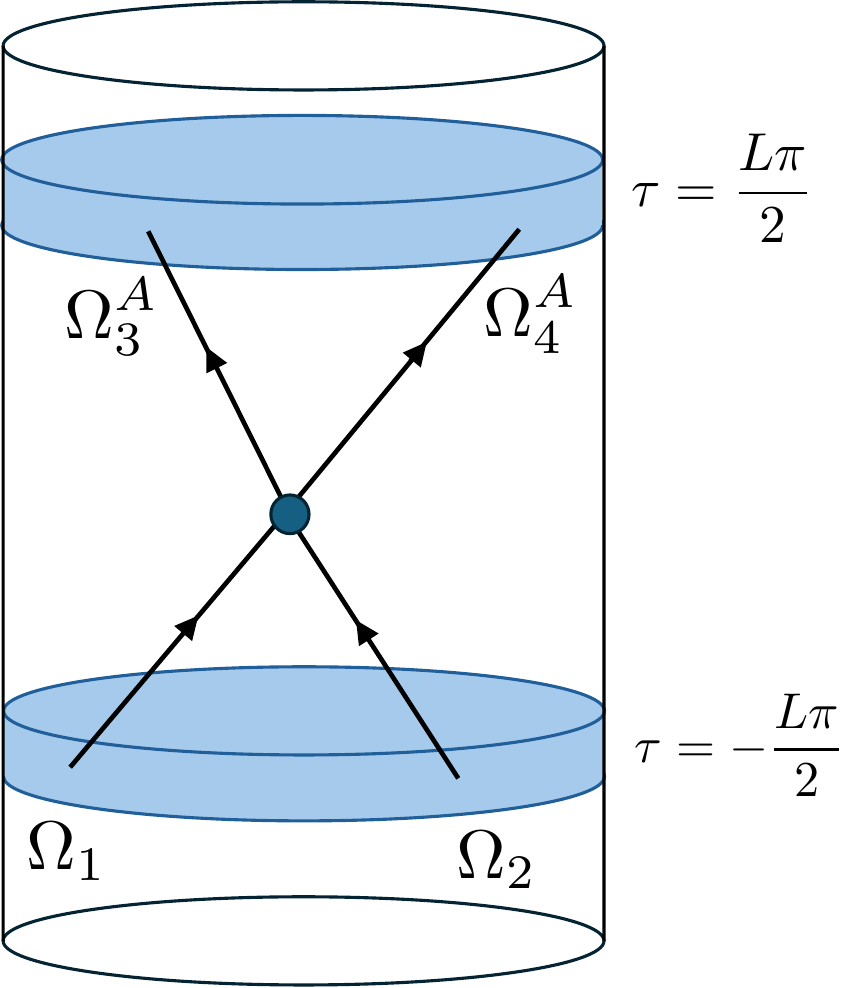}
\caption{\label{F:AdS} A cartoon of how a flat space scattering amplitude is obtained from the large radius limit of AdS/CFT.}
\end{center}
\end{figure}

\emph{Non-locality from the flat space limit of AdS/CFT.}
The source of the non-locality in the four-point function~\eqref{eq:extransf} can be traced to the interaction terms in the second and third lines of the commutator of boundary operators in~\eqref{eq:outincomm}. We will now see how this is consistent with the flat space limit of AdS/CFT~\cite{Giddings:1999jq,Penedones:2010ue,Fitzpatrick:2011jn,Hijano:2019qmi,Hijano:2020szl,Li:2021snj}. Consider a bulk theory that arises from the large radius limit of a theory in AdS, and work in coordinates where the global AdS line element is
\begin{equation}
	ds^2 = \frac{L^2}{\cos^2(\rho)}\left( -\frac{d\tau^2}{L^2} + d\rho^2 + \sin^2(\rho) d\Omega^2\right)\,,
\end{equation}
with $L$ the AdS radius, so that boundary time is $\tau$. The operators encoding the flat space $S$-matrix in the deep interior are smeared over time bands of $\mathcal{O}(1)$ width located at $\tau = \pm \frac{\pi L}{2}$. The operators encoding out-states are inserted in the band at $\tau = + \frac{L\pi}{2}$, whereas the operators encoding in-states are in the band at $\tau = -\frac{L\pi}{2}$. Moreover there is a relative antipodal flip between past and future, in the sense that in-states of definite spatial momentum are inserted at an angle $\Omega$ on the boundary sphere, and out-states of the same spatial momentum are inserted at the antipode $\Omega_A$. See Fig.~\ref{F:AdS}. One can see this antipodal flip at the level of symmetry algebra when you restrict the CFT isometries to the time bands around $\tau = \pm \frac{L\pi}{2}$ \cite{deGioia:2023cbd} and this antipodal flip is crucial to obtain Carrollian correlators in the limit \cite{Bagchi:2023fbj}.

If one considers a free theory in the bulk AdS, then of course one has the standard mode decomposition \cite{Balasubramanian:1999ri,Giddings:1999qu}. One can take the boundary limit and obtain the CFT operators in terms of those modes. If the bulk Hamiltonian is free, bulk time evolution evolves the modes at $\tau =-\frac{L\pi}{2}$ to antipodally-flipped ones at $\tau = \frac{L\pi}{2}$, so that the commutator of past and future modes is the standard equal time commutation relation. However, bulk interactions lead to corrections. The resulting evolution operator can still be perturbatively corrected as long as the bulk interactions obey microcausality~\cite{Fitzpatrick:2010zm}. Such bulk interactions result in anomalous dimensions for the boundary operators $O(\tau,\Omega)$ dual to the bulk field but nevertheless 
\begin{equation}
               \left[ O \!\left(\tau= +\frac{L\pi}{2},\Omega_{1A} \right)\!,\, O\! \left(\tau= -\frac{L\pi}{2},\Omega_2 \right) \right]
\end{equation}
will be proportional to \eqref{eq:outincomm}. From this perspective, the non-locality of Carrollian correlators simply arises because future operators are related to the past ones by evolution by the ADM Hamiltonian of global AdS.

\emph{A modified dictionary.}
The dictionary suggested by \eqref{eq:themap} generically implies that the codimension-one boundary description of $S$-matrices is non-local. Intuitively, this is expected from the fact that the degrees of freedom that encode scattering are respectively supported in $\mathscr{I}^-$ and $\mathscr{I}^+$ with antipodal matching conditions across $i^0$ \cite{Strominger:2013jfa}. \eqref{eq:extransf} embodies this simple fact. Under the current map \eqref{eq:themap}, the out- and in-operators become Carrollian fields at $\mathscr{I}^+$ and $\mathscr{I}^-$ respectively. Here, we postulate another dictionary that maps out- and in-operators to a single Carrollian field. However this is not enough to cure the non-locality uncovered above in \eqref{eq:extransf}.

Consider the boundary field built from both out- and in-oscillators,
\begin{equation}\label{eq:newmap1}
  \Psi(u,z,\bar z) \! = \!\!\int_0^{\infty}\!\!\!\! d\omega\, \omega^{\Delta\!-\!1} \! \big[ a_{\text{out}}(\omega,\!z,\!\bar z) e^{\!-i\omega u} \!- a^{\dagger}_{\text{in}}(\omega,\!z,\!\bar z) e^{i\omega u} \big] \, .
\end{equation}
Using the relation $a_{\text{out}}(\mathbf{p}) = \hat{S}^{\dagger} \, a_{\text{in}}(\mathbf{p}) \, \hat{S}$ where $\hat S$ is the $S$-matrix built from the interacting Hamiltonian,~\eqref{eq:newmap1} reduces to \eqref{eq:mapfields} for a free theory. In an interacting theory, however, the boundary field $\Psi(u,z,\bar z)$ encodes both the initial state (through $a_{\text{in}}$) and the full evolution (through $\hat S$). 
An analogous construction at $\mathscr{I}^-$ is
\begin{equation}
\begin{split}
\widetilde{\Psi}(v,z,\bar z) &= \int_0^{\infty}\! d\omega\, \omega^{\Delta-1} \big[\, a_{\text{in}}(\omega,z,\bar z)\, e^{-i\omega v}\\
& \qquad \qquad \qquad \quad - \hat S^{\dagger}\, a^{\dagger}_{\text{in}}(\omega,z,\bar z)\, \hat S e^{i\omega v} \big] \, .
\end{split}
\end{equation}
Antipodal matching, $\Psi(u,z,\bar z)\big|_{\mathscr{I}^+_-} =
  \widetilde{\Psi}(v,z,\bar z)\big|_{\mathscr{I}^-_+}$, would be a non-trivial requirement in the interacting theory, since $\Psi$ depends on the full history via $\hat S$. 

A transparent diagnostic of non-locality is
\begin{equation}
       \begin{split}
              &\big[\Psi(u_1,\!z_1,\!\bar z_1), \!\Psi(u_2,\!z_2,\!\bar z_2)\big] \!=\!\! \\ & \int \!\!d\omega_1 d\omega_2  \bigg( e^{\!-i\omega_1 \!u_1 + i\omega_2 u_2} \, \big[a_{\text{out}}(\omega_1,\!z_1,\!\bar z_1), a_{\text{in}}^{\dagger}(\omega_2,\!z_2,\!\bar z_2) \big] \label{eq:newmapcomm1} \\
& \!-\! e^{i\omega_1 \!u_1 \!- i\omega_2 u_2} \, \!\big[ a_{\text{out}}(\omega_2,\!z_2,\!\bar z_2), a_{\text{in}}^{\dagger}(\omega_1,\!z_1,\!\bar z_1) \big] \bigg)\,.
       \end{split}
\end{equation}
Using~\eqref{eq:outincomm} we find
\begin{widetext}
\begin{align}\label{eq:newmapcommfinal}
 & \big[\Psi(u_1,\!z_1,\!\bar z_1), \!\Psi(u_2,\!z_2,\!\bar z_2)\big] \!=\!\! \, (2\pi)^3\big(-i\pi\, \mathrm{sgn}(u_{12})\big)\, \delta^2(z_{12}) \\
 &\quad \quad \qquad \!+\! i\, \lambda \!\int d\omega_1\, d\omega_2\, d^4x \, \phi_I(x) \big( e^{i(p_1-p_2)\cdot x} e^{-i \omega_1 u_1 + i \omega_2 u_2}  - e^{i(p_2-p_1) \cdot x} e^{i \omega_1 u_1 - i \omega_2 u_2} \big)  \nonumber \\
 &\quad \quad \qquad \!+ \! \,\frac{\lambda^2}{12} \!\int d\omega_1\, d\omega_2\, d^4x_1\, d^4x_2\,\theta(x_1^0 - x_2^0)\,\Big( \!-\! 2\, \big( e^{i(p_1 - p_2)\cdot x_1} e^{-i \omega_1 u_1 +i \omega_2 u_2} - e^{i(p_2-p_1) \cdot x_1} e^{i \omega_1 u_1 - i \omega_2 u_2}  \big) \times  \nonumber \\
 &\quad \quad \qquad \times \big[\phi_I(x_1),\, \phi_I^3(x_2)\big]  +\, 6\, \big( e^{i p_1 \cdot x_1-ip_2\cdot x_2} e^{-i\omega_1 u_1 + i \omega_2 u_2} - e^{i p_2 \cdot x_1 - ip_1 \cdot x_2} e^{i \omega_1 u_1 -i \omega_2 u_2} \big)\, \big[\phi_I^2(x_1),\, \phi_I^2(x_2)\big] \Big) \, ,\nonumber 
\end{align}
\end{widetext}
where we have used $u_{12}\equiv u_1-u_2$ and $z_{12}\equiv z_1-z_2$ for brevity~\footnote{To keep the notation compact, we do not substitute the parametrization of the momenta $p_i$ from \eqref{eq:para}.}. Although the $\mathcal{O}(\lambda^2)$ term involves time ordering via $\theta(x_1^0-x_2^0)$, the spatial integrations extend over all space. The resulting operator insertions are therefore non-local, producing precisely the kind of correlations indicated by \eqref{eq:extransf}. In particular, the commutator does not vanish outside the Carrollian light-cone. 

\subsection*{Acknowledgements}

We thank A.~Bagchi, L.~Donnay, S.~Pasterski, R.~Ruzziconi and D.~Skinner for enlightening discussions. JC is supported by the Simons Collaboration on Celestial Holography, as well as a Fellowship from the Alfred P.~Sloan Foundation. PD and KJ are supported in part by an NSERC Discovery grant.

\appendix

\section{Regulating local \\ Carrollian field theories}

In this Appendix, we study local two-derivative Carrollian field theories. Carrollian field theories are sensitive both to the ultraviolet and the infrared, making their quantization subtle. In previous work \cite{Cotler:2024xhb}, two of us studied such models using a lattice regulator, with the result that many such models, including a Carrollian version of scalar QED, remain sensitive to the details of the lattice at long distance and thereby exhibit UV/IR mixing.

Since that work appeared there has been some discussion in the community of whether the results therein persist for other choices of regulator. Here we answer that question, using the suite of ``standard'' regulators to study Carrollian field theories. 

At the two derivative level, the basic species of Carrollian field theories are so-called ``electric theories'' like
\begin{align}
\label{E:electric0}
	S_E = \int du d^dx \left( \frac{1}{2}(\partial_u \phi)^2 - V(\phi)\right)\,,
\end{align}
with only time derivatives, ``magnetic theories'' like
\beq
	S_M = \int du d^dx \left( Q\partial_u \chi - \mathcal{H}(\chi,\nabla \chi)\right)\,,
\eeq
in which a field $\chi$ is constrained to be time-independent, or models which couple two such sectors together. Electric theories are sensitive to the ultraviolet thanks to their ultralocality, while magnetic theories are sensitive to the infrared thanks to the fact that $\chi$ is time-independent. With a lattice regulator, coupled theories behave like electric ones when it comes to sensitivity to the UV and IR.

In each of the following we consider simple Carrollian field theories regulated in standard ways. 

\paragraph{Momentum cutoff.}

First, consider imposing a hard cutoff $\Lambda$ on spatial momenta. This most closely resembles the spatial lattice regulator considered in~\cite{Cotler:2024xhb}. For theories of matter such a regulator is perfectly alright, while for gauge theories extra care is required (in the form of non-gauge-invariant counterterms at the cutoff scale) to ensure the quantum theory is gauge-invariant. In any case, consider a potential $V = \frac{1}{2}m^2 \phi^2 + \frac{\lambda_3}{3!}\phi^3+\frac{\lambda_4}{4!} \phi^4$, so that the lattice regulated theory is that of an on-site quantum mechanics with a quartic potential at each site.

In the free theory the thermal partition function is, for $V$ the spatial volume,
\begin{align}
\begin{split}
\label{E:Z01}
	\ln Z_0& = - \frac{1}{2} \beta V \sum_n \int \frac{d^dk}{(2\pi)^d}\ln (\omega_n^2 + m^2) 
	\\
	& = -  V \Lambda^d\frac{ \text{vol}(\mathbb{B}^d)}{(2\pi)^d} \ln \left( 2 \sinh\left( \frac{\beta m}{2}\right)\right)\,,
\end{split}
\end{align}
where $\text{vol}(\mathbb{B}^d)$ is the volume of a unit ball. We can compare this with a hypercubic lattice approximation, where the integral over momenta becomes the inverse volume of a unit cell,
\beq
	\int\frac{d^dk}{(2\pi)^d}\to \frac{1}{a^d}\,,
\eeq
with $a$ the lattice spacing. In the lattice approximation we have $Z_0 = Z_{\rm on-site}^N$ with $Z_{\rm on-site} =  \frac{1}{2\sinh\left( \frac{\beta m}{2}\right)}$ and $N = V/a^d$ the number of lattice sites. With a momentum cutoff we have essentially the same result with an effective number of lattice sites $N_{\rm eff} = V/a_{\rm eff}^d$ with an effective inverse lattice volume
\beq
	\frac{1}{a_{\rm eff}^d}  = \Lambda^d\frac{\text{vol}(\mathbb{B}^d)}{(2\pi)^d}\,.
\eeq
We see that the momentum cutoff $\Lambda$ regulates like a lattice with $\Lambda \sim 1/a$ as one would expect on dimensional grounds. 

When considering interactions or the correlation functions of composite operators, we run into similar spatial loop integrals. The loop integrands of electric theories are independent of spatial momenta, and so in a lattice approximation, or with a momentum cutoff, each spatial loop integral becomes the factor $1/a^d$ or $1/a_{\rm eff}^d$, so this resemblance persists in perturbation theory.

For example, taking $\lambda_3^2 \sim \lambda_4$ and treating both as small we have (using the free-field propagator $\frac{i}{\omega^2-m^2}$)
\begin{widetext}
\beq
\label{E:prop0}
	\langle \phi(\omega,k)\phi(-\omega,-k)\rangle = \frac{i}{\omega^2-m^2}+  \int \frac{d\omega' d^dk'}{(2\pi)^{d+1}}\left( - \frac{\lambda_3^2}{2} \frac{1}{(\omega'^2 - m^2)((\omega-\omega')^2 - m^2} + i   \lambda_4  \frac{1}{(\omega'^2-m^2)}\right) + \hdots\,.
\eeq
\end{widetext}
The one-loop correction simply gives a result proportional to $\Lambda^d$, so that in order to have a finite result we must scale the cubic and quartic couplings as $\lambda_3 = \Lambda^{-d/2} \tilde{\lambda}_3$ and $\lambda_4 = \Lambda^{-d} \tilde{\lambda}_4$ with tilde'd couplings held fixed as we send $\Lambda\to\infty$. This is the same scaling identified in~\cite{Cotler:2024xhb} with a lattice regulator upon making the identification $\Lambda \sim 1/a$. 

We can also consider correlation functions of composite operators. Even in the free theory correlations of monomials $\phi^n$ are UV-divergent unless we rescale the operator appropriately. For example, the $\phi^2$ two-point function is
\begin{widetext}
\beq
\label{E:composite0}
	\langle \phi^2(\omega,k) \phi^2(-\omega,-k)\rangle = - \int \frac{d\omega'}{2\pi}\frac{ d^dk'}{(2\pi)^d} \frac{1}{(\omega'^2-m^2)((\omega-\omega')^2-m^2)} + \hdots\,,
\eeq
\end{widetext}
which goes as $\Lambda^d$. The rescaled operator $\Lambda^{-d/2} \phi^2$ then has finite correlations. This is the same scaling identified in~\cite{Cotler:2024xhb}, again with the replacement $\Lambda \sim 1/a$. 

It is straightforward to see that the scalings of couplings and operators above are needed in order to maintain perturbativity as the cutoff is sent to infinity.

In contrast, the IR divergences present for magnetic theories are not regulated by a lattice regulator or a momentum cutoff, because after integrating over the degree of freedom $Q$, there is a residual integral over time-independent configurations of $\chi$ with an effective action $-T \int d^dx \,\mathcal{H}(\chi,\nabla\chi)$ with $T$ the length of the time-interval. Whether with a lattice regulator or momentum cutoff, the Hilbert space is spanned by eigenstates of $\chi(x)$ at each $x$ (where $x$ labels lattice sites in a lattice approximation, or $\chi(x)$ is built from a sum over Fourier modes $\tilde{\chi}(k)$ with $|k|<\Lambda$ with a momentum cutoff), and the evolution operator is a field space delta function,
\beq
	\langle \chi| \mathcal{U}(T)|\chi'\rangle = \exp\left( - i T \int d^dx \,\mathcal{H}(\chi,\nabla\chi)\right) \delta[\chi-\chi']\,.
\eeq
The frequency loop integrals that appear in correlation functions of $\chi$ are IR-divergent, and those divergences can be regulated by turning on a small temperature and then taking the temperature to zero. 

Electromagnetic theories have additional divergences coming from the short-distance limit of magnetic propagators, for instance $\lim_{x\to 0}\langle \chi(u,x)\chi(0,0)\rangle =- \int \frac{d^dk}{(2\pi)^d} \frac{i}{k^2+M^2}$ where $\chi$ has a tree-level mass $M$. These loop integrals are of course also regulated by a lattice or a momentum cutoff, going as $a^{2-d}$ or $\Lambda^{d-2}$ respectively.

\paragraph{Pauli-Villars.}

From the structure above, it is clear that a Pauli-Villars-like regulator where we add a Grassmann-odd field of large mass $m_{\rm PV}^2 \gg m^2$ does not regulate the momentum loop integrals appearing in the electric theory. For example, consider the two-point function of the composite operator $\phi^2$ in the free theory. With a Pauli-Villars field we have
\begin{widetext}
\begin{align}
	\langle \phi^2(\omega,k)& \phi^2(-\omega,-k)\rangle= - \int \frac{d\omega'}{2\pi}\frac{ d^dk'}{(2\pi)^d} \left( \frac{1}{(\omega'^2-m^2)((\omega-\omega')^2-m^2)}-\frac{1}{(\omega'^2-m_{\rm PV}^2)((\omega-\omega')^2-m_{\rm PV}^2)}\right)\,,
\end{align}
\end{widetext}
and the momentum integral is not regulated on account of the ultralocality of the regulator field.

\paragraph{Dimensional regularization.}

A third standard regulator is to dimensionally continue the number of spatial dimensions, $d \to d-\epsilon$, while keeping the time direction one–dimensional. For the electric scalar theory~\eqref{E:electric0}, e.g.~with a potential $V(\phi) = \frac{1}{2} m^2 \phi^2  + \frac{\lambda_3}{3!}\,\phi^3  + \frac{\lambda_4}{4!}\,\phi^4$ as before, all of the UV–sensitive quantities we encountered with a hard momentum cutoff, namely the free partition function~\eqref{E:Z01}, the one–loop correction to the propagator~\eqref{E:prop0}, and the composite two–point function~\eqref{E:composite0}, contain spatial integrals of the form $\int d^d k\, (k^2)^n$ whose integrands are independent of $k$ apart from possible powers of $k^2$ coming from insertions of spatial derivatives in the vertices.

In dimensional regularization, we analytically continue these spatial integrals to $d-\epsilon$ dimensions. Since no mass scale or external spatial momentum appears in the integrand, they are scaleless, and the standard dimensional–regularization prescription sets
\begin{equation}
\int \frac{d^{d-\epsilon} k}{(2\pi)^{d-\epsilon}}\,(k^2)^n = 0, \qquad n = 0,1,2,\ldots\,.
\label{E:scalelessint}
\end{equation}
Applying~\eqref{E:scalelessint} to the spatial integral in the free partition function~\eqref{E:Z01} immediately gives
\begin{equation}
\ln Z_0^{(\text{dim-reg})} = 0
\end{equation}
up to an additive constant independent of $\beta$ and $m$. Likewise, the one–loop self–energy in~\eqref{E:prop0} and its higher–loop generalizations reduce to products of scaleless spatial integrals and hence vanish, so the exact two–point function coincides with the tree–level ultralocal propagator. The same reasoning applied to~\eqref{E:composite0} shows that connected correlators of composite operators built purely from $\phi$ are also proportional to scaleless integrals and are therefore set to zero.

Exactly the same structure appears in electric Carrollian gauge theories: in the electric sector, the tree–level propagators are independent of spatial momentum, while spatial derivatives enter only through interaction vertices. Loop diagrams built purely from these electric degrees of freedom again collapse to scaleless spatial integrals of the form~\eqref{E:scalelessint} and hence vanish in dimensional regularization. In this scheme the electric sector is thus perturbatively trivial: the UV–sensitive contributions to the partition function, loop corrections, and composite correlators that were non–trivial with a hard cutoff or spatial lattice are all discarded.

The fact that the logarithm of the partition function and composite operators vanish is particularly fatal, as those facts are inconsistent with a quantum mechanical interpretation. For example, the vanishing of $\ln Z$ implies that the would-be Hilbert space has a single state, which moreover has $E=0$, which of course is inconsistent with a non-trivial two-point function of $\phi$. 

\section{A comment about \\ entanglement entropy}

One of the interesting results in the ``Carrollian holography'' program is a match between the interval entanglement entropy of a putative Carrollian conformal field theory, really a certain continuation of a two-point function of twist operators, with a computation in 3d gravity. See e.g.~\cite{Bagchi:2014iea}. 

Insofar as this result concerns the Hilbert space interpretation of a putative Carrollian dual, we have two comments about it. First, the ingredients that go into the field theory version of this computation, the two-point function of the twist operator, and the stress tensor in the presence of the twist operators, are sensitive to the ultraviolet and infrared. Even in pure 3d flat space gravity, where there is a boundary graviton description~\cite{Cotler:2024cia}, the amplitudes of that model have enormous IR sensitivities owing to the infinite number of conservation laws, so that in the absence of rotation, amplitudes are proportional to field-space delta functions, roughly $\delta[P-P']$ with $P$ and $P'$ the supermomentum profiles of the in- and out-states. The computations of~\cite{Bagchi:2014iea} involve a ratio of sensitive amplitudes. Optimistically, these sensitivities cancel, but such a claim should be carefully demonstrated.

More crucially, the ``modular flow'' obtained from these works is spacelike. See especially \cite{Jiang:2017ecm,Apolo:2020qjm,Apolo:2020bld}, e.g.~Eqs.~(3.27) and (3.28) and Fig.~2 of~\cite{Apolo:2020qjm}. In particular, the ``modular flow'' is tangent to the interval. This precludes any interpretation of these computations in terms of entanglement, reduced density matrices, and so on, even beyond the usual divergences that afflict position-space entanglement in quantum field theory.  The reason is that the putative ``reduced density matrix,'' the operator that enacts the flow, would for an interval at rest be $\exp\left( \int_0^{2\pi} d\theta\,f(\theta)J(\theta)\right)$ for $J(\theta)$ the super-angular momentum. This is not even an operator on account of $J$ being unbounded above and below, and in relativistic quantum field theory would be akin to the formal expression $\exp\!\left( \int dx \,u^{\mu}(x) P_{\mu}(x)\right)$ for $P_{\mu}$ the momentum operator, but for $u$ spacelike.



\bibliography{refs}
\bibliographystyle{JHEP}

\end{document}